\documentclass[runningheads]{llncs}
\pdfoutput=1

\newif\ifdraft\drafttrue
\newif\ifinlineref\inlinereffalse
\newif\iffinal\finalfalse
\newif\ifextended\extendedfalse
\newif\ifdotikz\dotikzfalse

\inlinereftrue
\draftfalse
\finaltrue

\def\papertitle{Eliminating Unfounded Set Checking for \hex-Programs}

\usepackage{times}
\usepackage[scaled]{helvet}
\usepackage{courier}

\usepackage{latexsym}
\usepackage{amsmath}
\usepackage{amssymb}
\usepackage{harpoon}

\usepackage{paralist}
\usepackage{multirow}

\usepackage{url}
\usepackage{color}

\iffinal\else
\fi
\ifdraft
\newcommand{\comment}[1]{{\bf\color{blue}{*** #1 ***}}}
\else
\newcommand{\comment}[1]{}
\fi

\usepackage{paralist}
\usepackage{enumerate}
\usepackage{booktabs}
\usepackage{alltt}
\usepackage{caption}

\usepackage{graphicx}
\graphicspath{{figures/}}

\usepackage{ifpdf}
\ifpdf
\usepackage{microtype}
\fi

\ifdraft
\usepackage{pdfsync}
\usepackage{srcltx}
\else
\fi

\ifdotikz
\usepackage{tikz}
\pgfrealjobname{unfoundedsetcheckaddon}
\usetikzlibrary{shapes,arrows,backgrounds,chains,%
matrix,patterns,arrows,decorations.pathmorphing,decorations.pathreplacing,%
positioning,fit,calc,decorations.text,shadows%
}
\else
\long\def\beginpgfgraphicnamed#1#2\endpgfgraphicnamed{\includegraphics{#1}}
\fi

\newcommand{\leanparagraph}[1]{\smallskip\noindent\textbf{#1}. }

\newenvironment{myitemize}{\begin{list}{$\bullet$}{%
\setlength{\topsep}{0pt}
\setlength{\leftmargin}{0pt}
\setlength{\itemindent}{10pt}}
\parskip=0pt
}
{\end{list}}

\renewcommand{\vec}[1]{{\bf #1}}

\newcommand{\ext}[3]{\ensuremath{\amp{#1}[#2](#3)}}
\DeclareMathOperator{\naf}{not}

\newcommand{\extfun}[1]{\ensuremath{f_{\text{\sl\&}#1}}}

\newcommand{\extsem}[4]{\ensuremath{f_{\text{\sl\&}#1}(#2,#3,#4)}}

\newcommand{\amp}[1]{\ensuremath{\text{\textsl{{\&}}}\!\mathit{#1}}}

\newcommand{\nop}[1]{}

\newcommand\hex{{\sc hex}}
\newcommand\dlvhex{{\sc dlvhex}}

\newcommand{\dlv}[0]{\texttt{DLV}}
\newcommand{\clasp}[0]{{\sc clasp}}

\newcommand{\smodels}[0]{{\sc smodels}}

\newcommand{\T}{\mathbf{T}}
\newcommand{\F}{\mathbf{F}}

\newcommand{\Assignment}{\ensuremath{\mathbf{A}}}
\newcommand{\AssignmentP}{\ensuremath{\hat{\mathbf{A}}}}

\newcommand{\Program}{\ensuremath{\Pi}}

\newcommand{\ProgramP}{\ensuremath{\hat{\Pi}}}

\newcommand{\overwrite}{\ensuremath{\stackrel{.}{\cup} \neg .}}

\newcommand\eqs{\ensuremath{\,{=}\,}}
\newcommand\ins{\ensuremath{\,{\in}\,}}
\newcommand\notins{\ensuremath{\,{\notin}\,}}
\newcommand\supsets{\ensuremath{\,{\supset}\,}}

\newcommand{\BIGOP}[1]
{\mathop{\mathchoice%
{\raise-0.22em\hbox{\huge $#1$}}%
{\raise-0.05em\hbox{\Large $#1$}}{\hbox{\large $#1$}}{#1}}}

\usepackage{ntheorem}

{\theorembodyfont{\normalfont}
\theoremheaderfont{\itshape}
\newtheorem*{proofsketch}{Proof (Sketch).}
}

\title{\papertitle%
\iffinal%
\thanks{This research has been supported by the Austrian Science
    Fund (FWF) project P20840, P20841, P24090, and by the Vienna Science and Technology
    Fund (WWTF) project ICT08-020.}
\fi
}

\author{Thomas Eiter \and Michael Fink \and Thomas Krennwallner \and
  Christoph Redl \and Peter Sch\"{u}ller}
\institute{
Institut f\"ur Informationssysteme, Technische Universit\"at Wien\\
\iffinal%
Favoritenstra\ss{}e\ 9-11, A-1040 Vienna, Austria\\
\fi
\email{$\{$eiter,fink,tkren,redl,ps$\}$@kr.tuwien.ac.at}
}

\newcommand\myfigureargu[1]{%
\begin{table}[#1]
  \footnotesize
  \renewcommand{\arraystretch}{1.15}
  \begin{center}
    \begin{tabular}[t]{r|r@{\,}r@{~}|r@{~~}r@{~}|r@{~}||r@{\,}r@{~}|r@{~~}r@{~}|r}
    \cline{1-11}
    \multirow{3}{*}{\rotatebox{90}{\#args}}
    & \multicolumn{5}{c||}{\raisebox{0ex}[2.5ex]{first answer set}}
    & \multicolumn{5}{c}{all answer sets} \\
    \cline{2-11}
    & \multicolumn{2}{c|}{\raisebox{0ex}[2.5ex]{standard approach}}
    & \multicolumn{2}{c|}{new approach}
    &
    & \multicolumn{2}{c|}{standard approach}
    & \multicolumn{2}{c|}{new approach}
    & \\
    & timeouts & avg & timeouts & avg & gain & timeouts & avg & timeouts & avg & gain \\
    \hline
5 & 0	& 1.09	& 0	& 1.07	& 2.16\%	& 0	& 1.70	& 0	& 1.56	& 8.44\% \\
6 & 0	& 2.40	& 0	& 2.30	& 4.38\%	& 0	& 4.58	& 0	& 3.74	& 18.42\% \\
7 & 0	& 5.58	& 0	& 5.33	& 4.47\%	& 0	& 15.66	& 0	& 11.28	& 27.95\% \\
8 & 0	& 14.26	& 0	& 12.74	& 10.70\%	& 3	& 71.06	& 2	& 39.32	& 44.66\% \\
9 & 0	& 39.82	& 0	& 33.57	& 15.70\%	& 16	& 174.99 & 8	& 106.34 & 39.23\% \\
10 & 2	& 126.54 & 0	& 80.00	& 36.78\%	& 40	& 278.98 & 16	& 214.81 & 23.00\% \\
    \hline
  \end{tabular}%
  \end{center}
  \caption{Argumentation Benchmarks: standard approach means the state-of-the-art approach without decomposition of the UFS check and without elimination of unnecessary checks, times are in seconds, timeout was 300 sec, for each system size there were 50 instances. }
  \label{fig:argumentation}
\end{table}
}%

\titlerunning{Eliminating Unfounded Set Checking for \hex-Programs}
\authorrunning{T.~Eiter \emph{et al}\/.}

\begin{document}

\setcounter{page}{83}

\setlength{\abovedisplayskip}{2pt}
\setlength{\abovedisplayshortskip}{2pt}
\setlength{\belowdisplayskip}{2pt}
\setlength{\belowdisplayshortskip}{2pt}

\maketitle

\begin{abstract}
\hex{}-programs are an extension of the Answer Set Programming (ASP) paradigm
incorporating external means of computation into the declarative programming
language through so-called external atoms. Their semantics is defined in terms of
minimal models of the Faber-Leone-Pfeifer (FLP) reduct.
Developing native solvers for \hex{}-programs based on an appropriate
notion of unfounded sets has been subject to recent research
for reasons of efficiency. Although this has lead to an improvement over naive
minimality checking using the FLP reduct, testing for foundedness remains
a computationally expensive task. In this work we improve on \hex{}-program
evaluation in this respect by identifying a syntactic class of programs,
that can be efficiently recognized and allows to entirely skip the
foundedness check. Moreover, we develop criteria for decomposing a program into
components, such that the search for unfounded sets can be restricted.
Observing that our results apply to many \hex{}-program applications provides
analytic evidence for the significance and effectiveness of our approach,
which is complemented by a brief discussion of preliminary experimental
validation.
\end{abstract}
\begin{keywords}
 Answer Set Programming, Nonmonotonic Reasoning, Unfounded Sets, FLP Semantics
\end{keywords}

\section{Introduction}
\label{sec:introduction}

In the last years, Answer Set Programming (ASP) has emerged as an
increasingly popular approach to declarative problem solving for a range
of applications \cite{brew-etal-11-asp}, thanks to expressive and
efficient systems like \smodels{}~\cite{simo-etal-2002},
\dlv{}~\cite{leon-etal-06-dlv}, cmodels~\cite{glm2006-jar}, and
\clasp{}~\cite{gks2012-aij}. However, recent developments in computing,
in which context awareness, distribution and heterogeneous information
sources gain importance, raised the need for access to external sources
in programs, be it in the context of the Web to access web services,
databases, or ontological information in different formats, in the
context of agents to acquire sensor input, etc.

To cater for this need, \hex{}-programs~\cite{eist2005} extend ASP
with so called external atoms, through which the user can couple any
external data source with a logic
program.  Roughly, such atoms pass information from the program, given
by predicate extensions, into an external source which
returns output values of an (abstract) function that it computes.
This extension has been utilized for a range of applications,
including querying data and ontologies on the Web,
multi-context reasoning, and reasoning about actions and planning, to mention a few
(cf.\ \cite{efiks2011-lpnmr}).
Notably, recursive data exchange between the rules and the external
sources is supported, which makes the formalism powerful.

The semantics of a \hex{}-program $\Program$ is
defined in terms of answer sets based on the  FLP reduct~\cite{flp2011-ai}:
an interpretation $\Assignment$ is an answer set of $\Program$, if and
only if it
is a~$\subseteq$-minimal model of the FLP-reduct $f\Program^{\Assignment}$ of $\Program$ wrt.\ $\Assignment$,
which is the set of all rules whose body is satisfied by $\Assignment$.
For ordinary logic programs, this semantics coincides with the one
where the canonical  GL-reduct \cite{gelf-lifs-91} is in place of
$f\Program^{\Assignment}$, and it is more appealing for extensions
with nonmonotonic aggregates \cite{flp2011-ai}, and the more general external atoms in \hex-programs.

The evaluation of a \hex-program $\Program$ in the \dlvhex{}\footnote{\url{http://www.kr.tuwien.ac.at/research/systems/dlvhex/}} solver
proceeds in two steps as
follows. In Step~1, external atoms are
viewed as ordinary atoms (\emph{replacement atoms}) and their truth
values are guessed by choice rules that are added. The resulting ordinary ASP program $\ProgramP$ is
then evaluated by an ordinary ASP solver and each of its answer sets $\AssignmentP$
is checked against the external sources, i.e., the guess is
verified. After that, the guess
for the non-replacement atoms, called $\Assignment$, is known to be a
model of $\Program$, and thus also of the reduct $f\Program^{\Assignment}$.  Step~2 then checks whether $\Assignment$ is
a $\subseteq$-minimal model  or, equivalently,
whether $\Assignment$ is unfounded-free \cite{faber2005-lpnmr}, i.e.,
there exists no unfounded set (UFS) of $\Program$ wrt.~$\Assignment$.

Unfortunately, Step 2 is computationally expensive in general, and it is
intractable even for Horn programs with nonmonotonic external atoms of
polynomial complexity, as follows from results in \cite{flp2011-ai}.
It is thus worthwhile to be aware of cases where this test is tractable,
or even better, superfluous such that Step~2 can be skipped.

Motivated by this issue, we consider in this paper programs $\Program$
for which the result of Step~1 is a $\subseteq$-minimal model of the
reduct $f\Program^{\Assignment}$.
We provide a sound \emph{syntactic criterion}
for deciding whether  the minimality check is needed,
and in further elaboration, we describe how a program can be
\emph{decomposed} into program components such that unfoundedness checks
can be delegated to the components, and the necessity of Step~2 thus be
assessed on a finer-grained level.

More in detail, our main contributions are the following:

\begin{myitemize}
\item We present a \emph{syntactic decision criterion} which can be used
  to decide whether a program possibly has unfounded sets. If the result
  of this check is negative, then the computationally expensive search
  for unfounded sets can be skipped. The criterion is based on atom
  dependency and, loosely speaking states that there are no cyclic
  dependencies of ground atoms through external atoms. This criterion
  can be efficiently checked for a given ground \hex{}-program using
  standard methods, and in fact applies to a range of applications, in
  particular, for input-stratified programs, where external sources are
  accessed in a workflow to produce input for the next stage of
  computation. However, there are relevant applications of
  \hex{}-programs where cycles through external atoms are essential,
  e.g., in encodings of problems on multi-context systems \cite{be2007} or abstract
  argumentation systems ~\cite{dung1995-aij}, for which Step~2 cannot be skipped.

\item In further elaboration, we consider a \emph{decomposition} of
  a program $\Program$ into components based on the dependency graph
  that is induced by the program. We show that~$\Program$ has some
  unfounded set with respect to the candidate answer set $\Assignment$
  if and only if (at least) one of the components $\Program_C$ in the
  decomposition has some unfounded set wrt.\ $\Assignment$; note that
  computing the decomposition is efficiently possible, and thus does not
  incur a large overhead. This allows us to apply the decision
  criterion for the necessity of Step~2 efficiently on a more fine-grained level, and
  the search for unfounded sets can be guided to relevant parts of the
  program. In particular, for the \hex{}-encoding of a Dung-style
  argumentation semantics \cite{dung1995-aij} which we consider, the decomposition
  approach yields a considerable gain, as shown in a preliminary experimental evaluation.
\end{myitemize}

This paper complements recent work on unfoundedness
checking for \hex{}-programs in~\cite{efkrs2012-jelia,eite-etal-12a}, which is part
of a larger effort to provide efficient evaluation of \hex{}-programs,
based on new algorithms cf.\ \cite{efkr2012-tplp}. By their wide
applicability, our results are significant especially for many
potential applications in practice.

\section{Preliminaries}
\label{sec:preliminaries}

In this section, we
start with some basic definitions, and then
introduce syntax and semantics of \hex{}-programs
and the notion of unfounded sets we are going to use.

A  \emph{(signed) literal}\/ is a positive or a negative formula
$\T a$ resp.\ $\F a$, where $a$ is a ground atom of form $p(c_1, \dotsc, c_\ell)$,
with predicate $p$ and %
constants
$c_1, \dotsc, c_\ell$,
abbreviated %
$p(\vec{c})$.
For a literal $\sigma \eqs \T a$ or $\sigma \eqs \F a$,
let $\overline{\sigma}$ denote its
opposite,
i.e., $\overline{\T a} \eqs \F a$ and $\overline{\F a} \eqs \T a$.

An \emph{assignment}\/ %
is a consistent set of literals
$\T a$ or $\F a$, where $\T a$ expresses that~$a \ins \mathcal{A}$
and
$\F a$ that %
$a \notins \mathcal{A}$.
$\mathcal{A}$ is \emph{complete}, also called an \emph{interpretation\/},
if no assignment $\Assignment' \supsets \Assignment$ exists.
We denote by $\Assignment^{\T} \eqs \{ a \mid \T a \ins \Assignment \}$
and $\Assignment^{\F} \eqs \{ a \mid \F a \ins \Assignment \}$
the set of atoms that are true, resp.\ false in $\Assignment$, and
by $\mathit{ext}(q, \Assignment) =
\{ \vec{c} \mid \T q(\vec{c}) \ins \Assignment \}$
the extension of a predicate $q$.
Furthermore, ${\Assignment}|_{q}$ is the set of all literals over
atoms of form $q(\vec{c})$ in $\Assignment$. For a list $\vec{q} =
q_1,\dotsc,q_k$ of predicates we write $p \in \vec{q}$ iff $q_i = p$
for some $1 \le i \le k$, and let~${\Assignment}|_{\vec{q}}
= \bigcup_{j} {\Assignment}|_{q_j}$.

A \emph{nogood} is a set $\{ L_1, \dotsc, L_n \}$ of
literals
$L_i, 1 \le i \le n$.
An interpretation $\Assignment$ is a \emph{solution} to a nogood $\delta$
(resp.\ a set $\Delta$ of nogoods), iff $\delta \not\subseteq \Assignment$
(resp.\ $\delta \not \subseteq \Assignment$ for all $\delta \in \Delta$).

\subsection{\hex-Programs}

\hex-programs were introduced
in~\cite{eist2005} as a generalization of (disjunctive)
extended logic programs under the answer set
semantics~\cite{gelf-lifs-91};
for details and background see~\cite{eist2005}.

\leanparagraph{Syntax}
\hex-programs extend ordinary ASP programs by \emph{external atoms},
which enable a bidirectional interaction between a program
and external sources of computation.
External atoms have a list of input parameters (constants or predicate names)
and a list of output parameters. Informally,
to evaluate an external atom, the reasoner passes the constants and extensions of the predicates
in the input tuple to the external source
associated with the external atom.
The external source
computes %
output tuples which are matched
with the output list. More formally,
a \emph{ground external atom} is of the form
\begin{equation}
\label{grexat}
\ext{g}{\vec{p}}{\vec{c}},
\end{equation}
where
$\vec{p} = p_1, \dotsc, p_k$ are constant input parameters (predicate names or object constants),
and
$\vec{c} = c_1, \dotsc, c_l$ are
constant output terms.

Ground \hex-programs are then defined similar to ground ordinary ASP programs.

\begin{definition}[Ground \hex-programs]
A ground \hex-program consists of rules
\begin{equation}
\label{rule}
  a_1\lor\cdots\lor a_k \leftarrow b_1,\dotsc, b_m, \naf\, b_{m+1},
  \dotsc, \naf\, b_n \ ,
\end{equation}
where each $a_i$ is an (ordinary) ground atom $p(c_1,\dotsc,c_\ell)$
with constants $c_i$, $1 \le i \le \ell$,
each~$b_j$ is either an ordinary ground atom or a ground external atom,
and $k+n>0$.%
\footnote{
For simplicity, we do not formally introduce strong negation
but view, as customary, classical literals $\neg a$ as new atoms
together with a constraint $\leftarrow a, \neg a$.
}
\end{definition}

The \emph{head} of a rule $r$ is
$H(r) = \{a_1, \dotsc, a_n \}$ and
the \emph{body}
is $B(r) = \{b_1, \dotsc, b_m,$ $\naf\, b_{m+1}, \dotsc, \naf\, b_n\}$.
We call $b$ or $\naf b$ in a rule body a \emph{default literal};
$B^{+}(r) = \{b_1, \dotsc, b_m\}$ is the \textit{positive body},
$B^{-}(r) = \{b_{m+1}, \dotsc, b_n\}$ is the \textit{negative body}.
For a program $\Program$, let $A(\Program)$ be the set of all ordinary atoms occurring in $\Program$.

We also use non-ground programs.
However, as suitable safety conditions allow for
using a grounding procedure \cite{eist2006},
we limit our investigation to ground programs.

\leanparagraph{Semantics and Evaluation}
Intuitively, a ground external atom $\ext{g}{\vec{p}}{\vec{c}}$
is true, if the external source $\amp{g}$ yields output tuple $\vec{c}$
when evaluated with input $\vec{p}$.
Formally, the semantics of a ground external atom $\ext{g}{\vec{p}}{\vec{c}}$
wrt.~an interpretation $\Assignment$ is given by the value of a $1{+}k{+}l$-ary Boolean
\emph{oracle function} $\extfun{g}$ that is defined for all possible values
of $\Assignment$, $\vec{p}$ and $\vec{c}$,
where $k$ is the length of $\vec{p}$ and $l$ is the length of $\vec{c}$.  Thus,
$\ext{g}{\vec{p}}{\vec{c}}$ is true relative
to $\Assignment$ if and only if it holds that
$\extsem{g}{\Assignment}{\vec{p}}{\vec{c}} = 1$.
Satisfaction of ordinary  rules and ASP programs~\cite{gelf-lifs-91}
is then extended to
\hex-rules and programs in the obvious way, and
the notion of extension $\mathit{ext}(\cdot, \Assignment)$
for external predicates $\amp{g}$ with input lists $\vec{p}$
is naturally defined by
$\mathit{ext}(\amp{g}[\vec{p}], \Assignment) =
\{ \vec{c} \mid \extsem{g}{\Assignment}{\vec{p}}{\vec{c}} = 1\}$.

\begin{definition}[FLP-Reduct~\cite{flp2011-ai}]
\label{def:flpreduct}
For an interpretation $\Assignment$ over a program $\Program$, the \emph{FLP-reduct}
$f\Program^{\Assignment}$ of $\Program$ wrt.\ $\Assignment$ is the set
$\{ r \in \Program \mid \Assignment \models b, \mbox{ for all } b \in B(r) \}$
of all rules whose body is satisfied under $\Assignment$.
\end{definition}

An assignment $\Assignment_1$ is smaller or equal to another assignment $\Assignment_2$
wrt.~a program $\Program$, denoted $\Assignment_1 \leq_\Program \Assignment_2$ iff
$\{ \T a \in \Assignment_1^{\T} \mid a \in A(\Program) \} \subseteq \{ \T a \in \Assignment_2^{\T} \mid a \in A(\Program) \}$.
\begin{definition}[Answer Set]
\label{def:answerset}
An answer set of $\Program$ is a $\leq_\Program$-minimal (complete) model $\Assignment$ of
$f \Program^\Assignment$.
\end{definition}

Since interpretations (and thus answer sets, etc.) are complete assignments, slightly abusing
notation, we adopt the usual convention to uniquely identify them with the set of
all positive literals they contain.

\begin{example}
\label{ex:id}
Consider the program $\Program%
  = \{
  p %
  \leftarrow \ext{\mathit{id}}{p}{}
  \}$,
where $\ext{\mathit{id}}{p}{}$ is true iff~$p$ is true. Then
$\Program$ has the answer set $\Assignment_1 = \emptyset$,
which is indeed %
a $\leq_\Program$-minimal model of~$f \Program^{\Assignment_1} = \emptyset$.
\end{example}
The answer sets of a \hex-program $\Program$ are
determined by the \dlvhex{} solver using a transformation
to ordinary ASP programs as follows.
Each external atom
$\ext{g}{\vec{p}}{\vec{c}}$
in~$\Program$ is replaced by an ordinary ground \emph{external replacement atom}
$e_{\amp{g}[\vec{p}]}(\vec{c})$
and a rule~
$e_{\amp{g}[\vec{p}]}(\vec{c}) \vee \mathit{ne}_{\amp{g}[\vec{p}]}(\vec{c}) \leftarrow$
is added to the program. The answer sets of the resulting \emph{guessing program} $\ProgramP$
are determined by an ordinary ASP solver and projected
to non-replacement atoms.
However, the resulting
interpretations are not necessarily models
of $\Program$, as the value of
$\amp{g}[\vec{p}]$ under $f_{\amp{g}}$
can be different from the one of $e_{\amp{g}[\vec{p}]}(\vec{c})$.
Each answer set of $\ProgramP$ is thus merely a \emph{candidate} %
which must be checked against the external sources.
If no discrepancy is found, the model candidate is a
\emph{compatible set} of~$\Program$. More precisely,
\begin{definition}[Compatible Set]
\label{def:compatibleset}
A \emph{compatible set} of a program $\Program$
is an interpretation~$\AssignmentP$  such that
\begin{compactenum}[(i)]
\item\label{en:cs1} $\AssignmentP$ is an answer set \cite{gelf-lifs-91} of the \emph{guessing program}
$\ProgramP$, and
\item\label{en:cs2} $\extsem{g}{\AssignmentP}{\vec{p}}{\vec{c}} = 1$ iff
$\T e_{\amp{g}[\vec{p}]}(\vec{c}) \in \AssignmentP$ for
all external atoms $\amp{g}[\vec{p}](\vec{c})$ in $\Program$,
i.e. the guessed values coincide with the actual output
under the input from $\AssignmentP$.
\end{compactenum}
\end{definition}
The compatible sets of $\Pi$ %
include (modulo
$A(\Pi)$) all (FLP) answer sets.
For each answer set $\Assignment$ there is a compatible set
$\AssignmentP$ such that $\Assignment$ is the restriction of
$\AssignmentP$ to non-replacement atoms, but not vice versa.
To filter out the compatible sets which are not answer sets,
the current evaluation algorithm
proceeds as follows.
Each compatible set $\Assignment$ is fed to the minimality check,
which is realized as a search for unfounded sets.
This is justified by
the following
Definitions~\ref{def:unfoundedset}~and~\ref{def:unfoundedfree}
and Theorem~\ref{def:answersetsufs} from~\cite{efkrs2012-jelia}.
(These results lift unfounded sets for disjunctive logic programs
with arbitrary aggregates~\cite{faber2005-lpnmr} to \hex{}-programs.)

\begin{definition}[Unfounded Set~\cite{efkrs2012-jelia}]
\label{def:unfoundedset}
Given a program $\Pi$ and an interpretation $\Assignment$, let $X$
be any set of ordinary ground atoms appearing in $\Pi$. Then,
$X$ is an \emph{unfounded set for %
$\Assignment$} iff, for each rule $r$ having some atoms
from $X$ in the head,
at least one of the following conditions holds,
where $\Assignment \overwrite X = (\Assignment \setminus \{ \T a \mid a \in X\}) \cup \{ \F a \mid a \in X\}$:
\begin{compactenum}[(i)]
\item some literal of $B(r)$ is false wrt.~$\Assignment$,
\item some literal of $B(r)$ is false wrt.~$\Assignment \overwrite X$, or
\item some atom of $H(r) \setminus X$ is true wrt.~$\Assignment$.
\end{compactenum}
\end{definition}

\begin{definition}[Unfounded-free Interpretations~\cite{efkrs2012-jelia}]
\label{def:unfoundedfree}
An interpretation $\Assignment$ of a program~$\Program$ is \emph{unfounded-free}
iff $\Assignment^{\T} \cap X = \emptyset$, for all unfounded sets $X$ of $\Program$ wrt.~$\Assignment$.
\end{definition}

\begin{theorem}[Characterization of Answer Sets~\cite{efkrs2012-jelia}]
\label{def:answersetsufs}
A model $\Assignment$ of a program $\Program$ is an answer set iff
it is \emph{unfounded-free}.
\end{theorem}

\begin{example}[cont'd]
\label{ex:id-contd}
Reconsider the program $\Program = \{\, p \leftarrow \ext{\mathit{id}}{p}{}\,\}$
from above.
Then the corresponding guessing program is
$\ProgramP
  = \{
  p %
  \leftarrow e_{\amp{\mathit{id}}[p]}();
  e_{\amp{\mathit{id}}[p]}{} \vee \mathit{ne}_{\amp{\mathit{id}}[p]}{}
  \leftarrow
  \}$
and has the answer sets $\Assignment_1 = \emptyset$ and
$\Assignment_2 = \{\T p, \T  e_{\amp{\mathit{id}}[p]}{}\}$.
While $\Assignment_1$ does not intersect with any unfounded sets and
is thus also a $\leq_\Program$-minimal model of $f \Program^{\Assignment_1} = \emptyset$, $A_2$ intersects with the unfounded set $U = \{p\}$ and is not an answer set.
\end{example}

Our \hex{} implementation \dlvhex{} realizes
the search for unfounded sets as a separate search
problem using an encoding as a SAT instance. That is,
for a program $\Program$ and an interpretation $\Assignment$
we construct a set of nogoods $\Gamma_{\Program}^{\Assignment}$
such that its solutions contain representations of all unfounded sets
of $\Program$ wrt.~$\Assignment$. A (relatively simple) post-check
finds the unfounded sets among the solutions of $\Gamma_{\Program}^{\Assignment}$.

\section{Deciding the Necessity of the UFS Check}
\label{sec:decisioncriterion}

An alternative to the search for unfounded sets is an explicit construction
of the reduct and a search for smaller models.
However, it turned out that the minimality check based on unfounded sets
is more efficient. Nevertheless the computational costs are still high.
Moreover, during evaluation of $\ProgramP$ for computing the
compatible set $\AssignmentP$, the ordinary ASP solver has already
made an unfounded set check, and we can safely assume that
it is founded from its perspective. Hence, all remaining unfounded
sets which were not discovered by the ordinary ASP solver have to
involve external sources, as their behavior is not fully
captured by the ASP solver.

In this section we formalize these ideas and
define a decision criterion which allows us to decide whether
a further UFS check is necessary for a given program.
We eventually define a class of programs which does not require
an additional unfounded set check.
Intuitively, we show that
every unfounded
set that is not already detected during the construction of $\AssignmentP$
contains input atoms of external atoms which are involved in cycles.
If no such input atom exists in the program, then the UFS check
is superfluous.

Let us therefore start with a definition of atom dependency.
\begin{definition}[Atom Dependency]
\label{def:atomdependency}
For a ground program $\Program$, and ground atoms $p(\vec{c})$ and $q(\vec{d})$,
we say that
\begin{compactenum}[(i)]
\item $p(\vec{c})$ \emph{depends} on $q(\vec{d})$, denoted $p(\vec{c}) \rightarrow q(\vec{d})$,
	iff for some rule $r \in \Program$ we have $p(\vec{c}) \in H(r)$ and $q(\vec{d}) \in B^{+}(r)$;
\item $p(\vec{c})$ \emph{depends externally} on $q(\vec{d})$, denoted $p(\vec{c}) \rightarrow_e q(\vec{d})$,
	iff for some rule $r \in \Program$ we have $p(\vec{c}) \in H(r)$ and there is a $\amp{g}[q_1, \ldots, q_n](\vec{e}) \in B^{+}(r) \cup B^{-}(r)$ with $q_i = q$ for some $1 \le i \le n$.
\end{compactenum}
\end{definition}

In the following, we consider \emph{dependency graphs} $G^R_{\Program}$ for a ground program
$\Program$, where the set of vertices is the set of all ground atoms, and the set of edges
is given by a binary relation $R$ over ground atoms. If $R$ is not explicitly mentioned, then it
is assumed to consist of $\rightarrow \cup \rightarrow_e$, whose elemtents are also
called \emph{ordinary edges} and \emph{e-edges}, respectively.

The next definition and lemma allow to restrict our attention to the ``core'' of an unfounded set,
i.e., its most essential part. For our purpose, we can then focus on such cores, disregarding atoms
in a cut which is defined as follows.
\begin{definition}[Cut]
\label{def:cut}
Let $U$ be an unfounded set of $\Program$ wrt.~$\Assignment$.
A set of atoms $C \subseteq U$ is called a \emph{cut}, iff
\begin{compactenum}[(i)]
\item $b \not \rightarrow_e a$,  for all $a \in C$ and $b \in U$ ($C$
  has no incoming or internal e-edges), and
\item $b \not \rightarrow a$ and $a \not \rightarrow b$, for all $a \in C$ and $b \in U \setminus C$ (there are no ordinary edges between $C$ and $U \setminus C$).
\end{compactenum}
\end{definition}
\begin{example}
\label{ex:cut}
Consider the program $\Program$ given as the following set of rules
\begin{align*}
r &\leftarrow \ext{\mathit{id}}{r}{} \\
p &\leftarrow \ext{\mathit{id}}{r}{}\\
p &\leftarrow q\\
q &\leftarrow p
\end{align*}
We have $p \rightarrow q$, $q \rightarrow p$, $r \rightarrow_e r$ and $p \rightarrow_e r$.
Program $\Program$ has the unfounded set~$U = \{p,q,r\}$ wrt.~$\Assignment = \{ \T p, \T q, \T r\}$.
Observe that $C = \{p,q\}$ is a cut, and therefore we have that~$U \setminus C = \{r\}$
is an unfounded set of $\Program$ wrt.~$\Assignment$.
\end{example}

We first prove that cuts can be removed from unfounded sets and the
resulting set is still an unfounded set.
\begin{lemma}[Unfounded Set Reduction Lemma]
\label{lemma:ufsreduction}
Let $U$ be an unfounded set of $\Program$ wrt.~$\Assignment$, and let $C$ be a cut.
Then, $Y = U \setminus C$ is an unfounded set of $\Program$ wrt.~$\Assignment$.
\end{lemma}
\begin{proofsketch}
If $Y = \emptyset$, then the result holds trivially.
Otherwise,
let $r \in \Program$ with $H(r) \cap Y \not= \emptyset$. We show that one of the conditions
in Definition~\ref{def:unfoundedset} holds.
Observe that $H(r) \cap U \not= \emptyset$ because $U \supseteq Y$.
Since $U$ is an unfounded set of $\Program$ wrt.~$\Assignment$, either
\begin{compactenum}[(i)]
\item $\Assignment \not \models b$ for some $b \in B(r)$; or
\item $\Assignment \overwrite U \not \models b$ for some $b \in B(r)$; or
\item $\Assignment \models h$ for some $h \in H(r) \setminus U$
\end{compactenum}
If (i), then the condition also holds wrt.~$Y$.

If (ii), let $a \in H(r)$ such that $a \in Y$,
and $b \in B(r)$ such that
$\Assignment \overwrite U \not \models b$.
We make a case distinction: either $b$ is an ordinary literal or an external one.

If it is an ordinary default-negated atom $\naf c$,
then $\Assignment \overwrite U \not\models b$ implies $\T c \in \Assignment$
and $c \not\in U$, and therefore also $\Assignment \overwrite Y \not\models b$.
So assume $b$ is an ordinary atom.
If $b \not \in U$ then $\Assignment \not \models b$ and case (i) applies,
so assume $b \in U$.
Because $a \in H(r)$ and $b \in B(r)$,
we have $a \rightarrow b$ and therefore
either $a, b \in C$ or $a, b \in Y$ (because there are no ordinary
edges between $C$ and $Y$). But by assumption $a \in Y$, and
therefore $b \in Y$, hence $\Assignment \overwrite Y \not \models b$.

If $b$ is an external literal, then there is no $q \in U$ with $a \rightarrow_e q$
and $q \not \in Y$. Otherwise, this would imply $q \in C$
and $C$ would have an incoming e-edge,
which contradicts the assumption that $C$ is a cut.
Hence, for all $q \in U$ with $a \rightarrow_e q$,
also $q \in Y$, and therefore the truth value of $b$ under
$\Assignment \overwrite U$ and $\Assignment \overwrite Y$ is the same.
Hence $\Assignment \overwrite Y \not \models b$.

If (iii), then also
$\Assignment \models h$ for some $h \in H(r) \setminus Y$ because $Y \subseteq U$
and therefore $H(r) \setminus Y \supseteq H(r) \setminus U$.
\hfill$\Box$
\end{proofsketch}

Next we prove, intuitively, that for each unfounded set $U$
of $\Program$,
either the input to some external atom is unfounded itself, or $U$ is already detected
when $\ProgramP$ is evaluated.
\begin{lemma}[EA-Input Unfoundedness]
\label{lemma:inputunfoundedness}
Let $U$ be an unfounded set of $\Program$ wrt.~$\Assignment$.
If there are no $x, y \in U$ such that $x \rightarrow_e y$,
then $U$ is an unfounded set of $\ProgramP$ wrt.~$\AssignmentP$.
\end{lemma}
\begin{proofsketch}
If $U = \emptyset$, then the result holds trivially.
Otherwise, let $\hat{r} \in \ProgramP$ such that $H(\hat{r}) \cap U \not= \emptyset$.
Let $a \in H(\hat{r}) \cap U$.
Observe that $\hat{r}$ cannot be an external atom guessing rule because
$U$ contains only ordinary atoms.
We show that one of the conditions
in Definition~\ref{def:unfoundedset}
holds for $\hat{r}$ wrt.~$\AssignmentP$.

Because $\hat{r}$ is no external atom guessing rule,
there is a corresponding rule $r \in \Program$
containing external atoms in place of replacement atoms. Because $U$
is an unfounded set of $\Program$ and $H(r) = H(\hat{r})$, either:
\begin{compactenum}[(i)]
\item $\Assignment \not \models b$ for some $b \in B(r)$; or
\item $\Assignment \overwrite U \not \models b$ for some $b \in B(r)$; or
\item $\Assignment \models h$ for some $h \in H(r) \setminus U$
\end{compactenum}
If (i), let $b \in B(r)$ such that $\Assignment \not \models b$
and $\hat{b}$ the corresponding literal in $B(\hat{b})$
(which is the same if $b$ is ordinary and the corresponding replacement
literal if $b$ is external). Then
also $\AssignmentP \not \models \hat{b}$ because $\AssignmentP$ is compatible.

For (ii), we make a case distinction: either $b$ is ordinary or external.

If $b$ is ordinary, then $b \in B(\hat{r})$ and $\AssignmentP \overwrite U \not \models b$
holds because $\Assignment$ and $\AssignmentP$ are equivalent for ordinary atoms.

If $b$ is an external atom or default-negated external atom, then
no atom $p(\vec{c}) \in U$ is input to it, i.e. $p$ is
not a predicate input parameter of $b$; otherwise
we had $a \rightarrow_e p(\vec{c})$, contradicting our assumption
that $U$ has no internal e-edges.
But then $\Assignment \overwrite U$ implies $\Assignment \not \models b$
because the truth value of $b$ under $\Assignment \overwrite U$
and $\Assignment$ is the same. Therefore we can apply case (i).

If (iii), then also $\AssignmentP \models h$ for some $h \in H(\hat{r}) \setminus U$
because $H(r) = H(\hat{r})$ contains only ordinary atoms
and $\Assignment$ is equivalent to $\AssignmentP$
for ordinary atoms.
\hfill$\Box$
\end{proofsketch}
\begin{example}
\label{ex:eaunfoundedness}
Reconsider the program $\Program$ from Example~\ref{ex:cut}.
Then the unfounded set $U' = \{p,q\}$ wrt.~$\Assignment' = \{\T p, \T q, \F r\}$
is already detected when $\ProgramP$ consisting of
\begin{align*}
 e_{\amp{\mathit{id}}[r]}() \vee \mathit{ne}_{\amp{\mathit{id}}[r]}() &\leftarrow\\
 r &\leftarrow e_{\amp{\mathit{id}}[r]}()\\
 p &\leftarrow e_{\amp{\mathit{id}}[r]}()\\
 p &\leftarrow q\\
 q &\leftarrow p
\end{align*}
is evaluated by the ordinary ASP solver because $p \not\rightarrow_e q$ and $q \not\rightarrow_e p$.
In contrast, the unfounded set $U'' = \{p,q,r\}$ wrt.~$\Assignment'' = \{\T p, \T q, \T r\}$
is \emph{not} detected by the ordinary ASP solver because $p,r \in U''$ and $p \rightarrow_e r$.
\end{example}

The essential property of unfounded sets of $\Program$ wrt.~$\Assignment$ that are not recognized
during the evaluation of $\ProgramP$, is the existence of
cyclic dependencies including input atoms of some external atom.
Towards a formal characterization of a class of programs without this property, i.e.,
that do not require additional UFS checks, we define cycles as follows.
\begin{definition}[Cycle]
\label{def:cycle}
A \emph{cycle} under a binary relation $\circ$ is a sequence of elements $C = c_0, c_1, \ldots, c_n, c_{n+1}$
with $n \ge 0$,
such that $(c_i, c_{i+1}) \in \circ$ for all $0 \le i \le n$ and $c_0 = c_{n+1}$.
We say that a set $S$ \emph{contains a cycle} under $\circ$, if
there is a cycle $C = $ $c_0, c_1, \ldots, c_n, c_{n+1}$ under $\circ$ such that $c_i \in S$ for all $0 \le i \le n+1$.
\end{definition}

The following proposition states, intuitively, that each unfounded set
$U$ of $\Program$ wrt.~$\Assignment$
which contains no cycle through the input atoms to some external atom
has a corresponding unfounded set $U'$ of $\ProgramP$ wrt.~$\AssignmentP$.
That is, the unfoundedness is already detected when $\ProgramP$ is evaluated.

Let $\rightarrow^{d} \;=\; \rightarrow \cup \leftarrow \cup \rightarrow_e$,
where $\leftarrow$ is the inverse of $\rightarrow$, i.e. $\leftarrow
\;=\; \{ (x,y) \mid (y,x) \in\; \rightarrow \}$.
A cycle $c_0, c_1, \ldots, c_n, c_{n+1}$ under $\rightarrow^{d}$
is called an \emph{e-cycle}, iff it contains e-edges, i.e.,
iff $(c_i, c_{i+1}) \in \rightarrow_e$ for some $0 \le i \le n$.
\begin{proposition}[Relevance of e-cycles]
\label{prop:ecyclerelevance}
Let $U \not= \emptyset$ be an unfounded set of $\Program$ wrt.~$\Assignment$
that does not contain any e-cycle under $\rightarrow^{d}$.
Then, there exists a nonempty unfounded set of $\ProgramP$ wrt.~$\AssignmentP$.
\end{proposition}
\begin{proofsketch}
We define the \emph{reachable set} $R(a)$ from some atom $a$ as
$$R(a) = \{ b \mid (a,b) \in \{\rightarrow \cup \leftarrow\}^{*} \},$$
i.e. the set of atoms $b \in U$ reachable from $a$ using edges from $\rightarrow \cup \leftarrow$ only but no e-edges.

We first assume that $U$ contains at least one e-edge, i.e. there are $x, y \in U$ such that $x \rightarrow_e y$.
Now we show that there is a $u \in U$ with outgoing e-edge (i.e. $u \rightarrow_e v$ for some $v \in U$),
but such that $R(u)$ has no incoming e-edges (i.e.
for all $v \in R(u)$ and $b \in U$, $b \not \rightarrow_e v$ holds).
Suppose to the contrary that for all $a$ with outgoing e-edges,
the reachable set $R(a)$ has an incoming e-edge.
We now construct an e-cycle
under $\rightarrow^{d}$,
which contradicts our assumption.
Start with an arbitrary node with an outgoing e-edge $c_0 \in U$ and let $p_0$
be the (possibly empty) path (under $\rightarrow \cup \leftarrow$) from $c_0$ to the node
$d_0 \in R(c_0)$ such that $d_0$ has an incoming e-edge, i.e. there is a $c_1$ such that $c_1 \rightarrow_e d_0$;
note that $c_1 \not \in R(c_0)$\footnote{
Whenever $x \rightarrow_e y$ for $x, y \in U$,
then there is no path from $x$ to $y$ under $\rightarrow \cup \leftarrow$,
because otherwise we would have an e-cycle under $\rightarrow^{d}$.
}.
By assumption, also some node $d_1$ in $R(c_1)$ has an incoming e-edge (from some node $c_2 \not \in R(c_1)$). Let $p_1$ be the path from
$c_1$ to $d_1$, etc. By iteration we can construct the
concatenation of the paths $p_0, (d_0, c_1), p_1, (d_1, c_2), p_2, \ldots, p_i, (d_i, c_{i+1}), \ldots$,
where the $p_i$ from $c_i$ to $d_i$
are the paths within reachable sets, and the $(d_i, c_{i+1})$ are the e-edges between reachable sets.
However, as $U$ is finite some nodes on this path must be equal, i.e., a prefix of the
constructed sequence represents an e-cycle (in reverse order).

This proves that $u$ is a node with outgoing e-edge but such that $R(u)$ has no incoming e-edges.
We next show that $R(u)$ is a cut. Condition (i) is immediately satisfied
by definition of $u$. Condition (ii) is shown as follows.
Let $u' \in R(u)$ and $v' \in U \setminus R(u)$. We have to show that $u' \not \rightarrow v'$
and $v' \not \rightarrow u'$. Suppose, towards a contradiction, that $u' \rightarrow v'$.
Because of $u' \in R(u)$, there is a path from $u$ to $u'$ under $\rightarrow \cup \leftarrow$.
But if $u' \rightarrow v'$, then
there would also be a path from $u$ to $v'$ under $\rightarrow \cup \leftarrow$ and $v'$ would be in $R(u)$, a contradiction
Analogously, $v' \rightarrow u'$ would also imply that there is a path from $u$ to $v'$
because there is a path from $u$ to $u'$, again a contradiction.

Therefore, $R(u)$ is a cut of $U$, and by Lemma~\ref{lemma:ufsreduction}, it follows that $U \setminus R(u)$ is an unfounded set.
Observe that $U \setminus R(u)$ contains one e-edge less than $U$ because $u$ has an outgoing e-edge.
Further observe that $U \setminus R(u) \not= \emptyset$ because there is a $w \in U$
such that $u \rightarrow_e w$ but $w \not \in R(u)$.
By iterating this argument, the number of e-edges in the unfounded set can be reduced to zero in a
nonempty core.
Eventually, Lemma~\ref{lemma:inputunfoundedness} applies, proving that the remaining set
is an unfounded set of $\ProgramP$.
\hfill$\Box$
\end{proofsketch}
\begin{corollary}
\label{cor:ecyclefreeprograms}
If there is no e-cycle under $\rightarrow^{d}$
and $\ProgramP$ has no unfounded set wrt.~$\AssignmentP$, then~$\Assignment$ is unfounded-free for $\Program$.
\end{corollary}
\begin{proofsketch}
Suppose there is an unfounded set $U$ of $\Program$ wrt.~$\Assignment$.
Then it contains no e-cycle because there is no e-cycle under $\rightarrow^{d}$.
Then by Proposition~\ref{prop:ecyclerelevance} there is an unfounded set of $\ProgramP$
wrt.~$\AssignmentP$, which contradicts our assumption.
\hfill$\Box$
\end{proofsketch}

This corollary can be used as follows to increase performance of an evaluation algorithm:
if there is no cycle under $\rightarrow^{d}$ containing e-edges,
then an explicit unfounded set check is not necessary
because the unfounded set check made during evaluation of~$\ProgramP$
suffices.
Note that this test can be done efficiently (in fact in linear time, similar to
deciding stratifiability of an ordinary logic program).
Moreover, in practice one can abstract from
$\rightarrow^{d}$ by using analogous relations on the
level of predicate symbols instead of atoms. Clearly, if there is no e-cycle
in the predicate dependency graph,
then there can also be no e-cycle in the atom dependency graph.
Hence, the predicate dependency graph can be used to decide whether
the unfounded set check can be skipped.
\begin{example}
All example programs considered until here require an UFS check, but
the program $\Program = \{ \mathit{out}(X) \leftarrow \ext{\mathit{diff}}{\mathit{set}_1, \mathit{set}_2}{X} \} \cup F$
does not for any set of facts $F$, because there is no e-cycle under $\rightarrow^{d}$,
where $\mathit{diff}$ computes the set difference of the extensions of $\mathit{set}_1$ and $\mathit{set}_2$.

Also $\Program = \{ \mathit{str}(Z) \leftarrow \mathit{dom}(Z), \mathit{str}(X), \mathit{str}(Y), \naf \ext{\mathit{concat}}{X,Y}{Z} \}$
(where $\amp{\mathit{concat}}$ takes two constants and computes their string concatenation) does not need such a check;
there is a  cycle over an external atom, but no e-cycle under $\rightarrow^{d}$.
\end{example}

Moreover, the following proposition states that, intuitively, if $\ProgramP$
has no unfounded sets wrt.~$\AssignmentP$, then
any unfounded set $U$ of $\Program$ wrt.~$\Assignment$ must contain
an atom which is involved in a cycle under $\rightarrow^{d}$ that has an e-edge.
\begin{definition}[Cyclic Input Atoms]
For a program $\Program$, an atom $a$ is a \emph{cyclic input atom},
iff there is an atom $b$ such that $b \rightarrow_e a$ and there is
a path from $a$ to $b$ under $\rightarrow^{d}$.
\end{definition}

Let $\mathit{CA}(\Program)$ denote the set of
all cyclic input atoms of program $\Program$.
\begin{proposition}[Unfoundedness of Cyclic Input Atom]
\label{prop:unfoundednesscyclicpredicateinput}
Let $U\neq\emptyset$ be an unfounded set of $\Program$ wrt.~$\Assignment$ such that
 $U$ does not contain cyclic input atoms.
Then, $\ProgramP$ has a nonempty unfounded set wrt.~$\AssignmentP$.
\end{proposition}
\begin{proofsketch}
If $U$ contains no cyclic input atoms, then all
cycles under $\rightarrow^{d}$ containing e-edges
in the atom dependency graph of $\Program$
are broken, i.e. $U$
does not contain an e-cycle under $\rightarrow^{d}$.
Then by Proposition~\ref{prop:ecyclerelevance}
there is an unfounded set of $\ProgramP$ wrt.~$\AssignmentP$.
\hfill$\Box$
\end{proofsketch}

Proposition~\ref{prop:unfoundednesscyclicpredicateinput}
allows for generating the additional nogood
$\{ \F a \mid a \in \mathit{CA}(\Program) \}$
and adding it to $\Gamma^{\Assignment}_{\Program}$.
Again, considering predicate symbols instead of atoms is possible
to reduce the overhead introduced by the dependency graph.

\section{Program Decomposition}

It turns out that the usefulness of the decision criterion can be increased
by decomposing the program into components, such that
the criterion can be applied component-wise. This allows for restricting
the unfounded set check to components with e-cycles,
whereas e-cycle-free components can be ignored in the check.

Let $\mathcal{C}$ be a partitioning of the ordinary atoms $A(\Program)$
of $\Program$
into subset-maximal strongly connected components under $\rightarrow \cup \rightarrow_e$.
We define
for each partition $C \in \mathcal{C}$ the subprogram $\Program_{C}$
\emph{associated with $C$} as
$\Program_{C} = \{ r \in \Program \mid H(r) \cap C \not= \emptyset \}$.

We next show that if a program %
has an unfounded set $U$ wrt.~$\Assignment$,
 then $U\,{\cap}\,C$ is an unfounded set wrt.~$\Assignment$ for the subprogram of some strongly
connected component $C$.
\begin{proposition}
\label{prop:programdecomposition}
Let $U \not= \emptyset$ be an unfounded set of $\Program$ wrt.~$\Assignment$.
Then, for some $\Program_{C}$ with $C \in \mathcal{C}$ it holds
that $U \cap C$ is a nonempty unfounded set of $\Program_{C}$ wrt.~$\Assignment$.
\end{proposition}
\begin{proofsketch}
Let $U$ be a nonempty unfounded set of $\Program$ wrt.~$\Assignment$.
Because $\mathcal{C}$ is a decomposition of $A(\Program)$ into strongly
connected components, the \emph{component dependency graph}
$$\left\langle \mathcal{C},
\{ (C_1, C_2) \mid C_1, C_2 \in \mathcal{C}, \exists a_1 \in C_1, a_2 \in C_2:  (a_1, a_2) \in \rightarrow \cup \rightarrow_e \} \right\rangle$$
is acyclic.
Following the hierarchical component dependency graph from the
nodes without predecessor components downwards, we can find
a ``first'' component which has a nonempty intersection with $U$,
i.e., there exists a component $C \in \mathcal{C}$ such that
$C \cap U \not= \emptyset$ but $C' \cap U = \emptyset$ for all transitive
predecessor components $C'$ of $C$.

We show that $U \cap C$ is an unfounded set of $\Program_{C}$ wrt.~$\Assignment$.
Let $r \in \Program_{C}$ be a rule such that $H(r) \cap (U \cap C) \not= \emptyset$.
We have to show that one of the conditions of Definition~\ref{def:unfoundedset}
holds for $r$ wrt.~$\Assignment$ and $U \cap C$.

Because $U$ is an unfounded set of $\Program$ wrt.~$\Assignment$
and $H(r) \cap (U \cap C) \not= \emptyset$ implies $H(r) \cap U \not= \emptyset$, we know
that one of the conditions holds for $r$ wrt.~$\Assignment$ and $U$.
If this is condition (i) or (iii), then it trivially holds also wrt.~$\Assignment$
and $U \cap C$ because these conditions depend only on the assignment $\Assignment$, but not on
the unfounded set $U$.

If it is condition (ii), then $\Assignment \overwrite U \not\models b$
for some (ordinary or external) body literal $b \in B(r)$.
We show next that the truth value of all literals in
$B(r)$ is the same under $\Assignment \overwrite U$ and $\Assignment \overwrite  (U \cap C)$,
which proves that condition (ii) holds also wrt.~$\Assignment$ and~$U \cap C$.

If $b = \naf a$ for some atom $a$, then $\T a \in \Assignment$ and $a \not\in U$
and consequently $a \not\in U \cap C$, hence $\Assignment \overwrite (U \cap C) \not\models b$.
If $b$ is an ordinary atom, then either $\F b \in \Assignment$,
which implies immediatly that $\Assignment \overwrite (U \cap C) \not\models b$,
or $b \in U$. But in the latter case $b$ is either
in a predecessor component $C'$ of $C$ or in $C$ itself
(since $h \rightarrow b$ for all $h \in H(r)$).
But since $U \cap C' = \emptyset$
for all predecessor components of $C$, we know $b \in C$ and therefore $b \in (U \cap C)$,
which implies $\Assignment \overwrite (U \cap C) \not\models b$.

If $b$ is a positive or default-negated external atom,
then all input atoms $a$ to $b$ are either
in a predecessor component $C'$ of $C$ or in $C$ itself
(since $h \rightarrow_e a$ for all $h \in H(r)$).
We show with a similar argument as before that the truth value of each
input atom $a$ is the same under $\Assignment \overwrite U$
and $\Assignment \overwrite (U \cap C)$:
if $\Assignment \overwrite U \models a$, then $\T a \in \Assignment$
and $a \not\in U$, hence $a \not\in (U \cap C)$ and
therefore $\Assignment \overwrite (U \cap C) \models a$.
If $\Assignment \overwrite U \not\models a$, then either
$\F a \in \Assignment$, which immediately implies
$\Assignment \overwrite (U \cap C) \not\models a$,
or $a \in U$.
But in the latter case $a$ must be in $C$ because $U \cap C' = \emptyset$
for all predecessor components $C'$ of $C$. Therefore $a \in (U \cap C)$
and consequently $\Assignment \overwrite (U \cap C) \not\models a$.
Because all input atoms $a$ have the same truth value under
$\Assignment \overwrite U$ and $\Assignment \overwrite (U \cap C)$,
the same holds also for the positive or default-negated external atom $b$
itself.
\hfill $\Box$
\end{proofsketch}

This proposition states that a search for unfounded sets can be done
independently for the subprograms $\Program_{C}$
for all $C \in \mathcal{C}$.
If there exists a global unfounded set, then there exists also one
in at least one of the program components.
However, we know by Corollary~\ref{cor:ecyclefreeprograms} that
programs $\Program$ without e-cycles cannot contain unfounded sets, which are not
already detected when $\ProgramP$ is solved. If we apply this proposition
to the subprograms $\Program_{C}$, we can safely ignore e-cycle-free
program components.
\begin{example}
Reconsider the program $\Program$ from Example~\ref{ex:cut}.
Then $\mathcal{C}$ contains the components $C_1 = \{p,q\}$
and $C_2 = \{r\}$ and we have
$\Program_{C_1} = \{ p \leftarrow \amp{\mathit{id}[r]}(); p \leftarrow q; q \leftarrow p \}$
and
$\Program_{C_2} = \{ r \leftarrow \amp{\mathit{id}[r]}() \}$.
By Proposition~\ref{prop:programdecomposition}, each unfounded set
of $\Program$ wrt.~some assignment can also detected over one of the components.
Consider e.g. $U = \{p,q,r\}$ wrt.~$\Assignment = \{ \T p, \T q, \T r\}$.
Then $U \cap \{r\} = \{r\}$ is also an unfounded set of $\Program_{C_2}$ wrt.~$\Assignment$.

By separate application of Corollary~\ref{cor:ecyclefreeprograms} to the
components, we can conclude that there can be no unfounded sets over $\Program_{C_1}$
that are not already detected when $\ProgramP$ is evaluated (because it has no e-cycles).
Hence, the additional unfounded set check is only necessary for $\Program_{C_2}$.
Indeed, the only unfounded set which is not detected when $\ProgramP$ is
evaluated is $\{r\}$ of $\Program_{C_2}$ wrt.~any interpretation $\Assignment \supseteq \{\T r\}$.
\end{example}

Finally, one can also show that splitting, i.e., the component-wise check for foundedness,
does not lead to spurious unfounded sets.
\begin{proposition}
If $U$ is an unfounded set of $\Program_{C}$
wrt.~$\Assignment$ such that $U \subseteq C$, then $U$
is  an unfounded set of $\Program$ wrt.~$\Assignment$.
\end{proposition}
\begin{proofsketch}
If $U = \emptyset$, then the result holds trivially.
By definition of $\Program_{C}$ we have $H(r) \cap C = \emptyset$ for all
$r \in \Program \setminus \Program_{C}$.
By precondition of the proposition we have $U \subseteq C$.
But then $H(r) \cap U = \emptyset$ for all $r \in \Program \setminus \Program_{C}$
and $U$ is an unfounded set of $\Program$ wrt.~$\Assignment$.
\hfill $\Box$
\end{proofsketch}

\section{Implementation and Evaluation}
\label{sec:implementation}

\myfigureargu{tb}

For implementing our technique,
we integrated \clasp{} into our prototype system \dlvhex{};
we use \clasp{} as an ASP solver for computing compatible sets
and as a SAT solver for solving the nogood set of the UFS check.
We evaluated the implementation
on a Linux server with two 12-core AMD 6176 SE CPUs with 128GB RAM.

\leanparagraph{Argumentation Benchmarks}
In this benchmark we compute
ideal set extensions for randomized instances
of abstract argumentation frameworks~\cite{dung1995-aij}
of different sizes.
In these instances, the cycles involve usually
only small parts of the overall programs, hence
the program decomposition is very effective.
Table~\ref{fig:argumentation} shows results of our experimental evaluation
on argumentation benchmark instances;
for computing average times, we considered 300 seconds for instances that timed out.
The encodings contain a cyclic part with cycles over external atoms,
and a cyclic part with cycles that do not contain external atoms.
Therefore in these instances our new approach can help in limiting the set of atoms
for which unfounded sets must be checked,
which explains the significant performance gain
due to less time spent in the UFS check.

\leanparagraph{Multi-Context System Benchmarks}
MCSs~\cite{be2007} are a formalism for interlinking knowledge based systems;
in~\cite{efsw2010-kr}, \emph{inconsistency explanations (IEs)}\/ for an MCS
were defined.
This benchmark
computes the IEs, which correspond 1-1 to answer sets of an encoding
rich in cycles through external atoms
(which evaluate local knowledge base semantics).
We use random instances of different topologies
created with an available benchmark generator.
For the MCS benchmarks we tested 68 consistent and 88 inconsistent
MCSs for which we compute inconsistency explanations~\cite{efsw2010-kr}.
This encoding contains saturation over external atoms,
where nearly all cycles in the \hex{}-program
contain at least one external atom.
Therefore the methods we introduce in this work
can only very rarely reduce the set of atoms
for which the UFS check needs to be performed.

The benchmark result for MCS instances
confirms that the syntactic check we introduce in this paper
is very cheap and does not impede %
performance,
even if an instance does not admit a considerable simplification for
the UFS check:
over all 156 instances,
we had an overall runtime of 25357 seconds
with the standard approach,
and a runtime of 25115 seconds with our new approach;
the gain is 242 seconds which is less than one percent speedup
(for enumerating all inconsistency explanations) by applying our method.
This is a very small gain,
and there is no difference in the number of instances that timed out.

\leanparagraph{Default Reasoning over Description Logics Benchmarks}
Another application of \hex-programs is the
DL-plugin~\cite{eiks2009-amai}, which integrates description logics
ontologies with rules. This allows, for instance, default reasoning over description
logic knowledge bases, which is not possible in DL knowledge bases alone.
Defaults require cyclic dependencies over external atoms.
However, as all such dependencies involve default negated atoms,
we have no cycles according to Definition~\ref{def:atomdependency},
which respects only positive dependencies. Hence, the decision criterion
comes to the conclusion that no UFS check is required.

We used variants of the benchmarks presented in~\cite{efkr2012-tplp},
which query wines from an ontology and
classify them as red or white wines, where a wine is assumed to be white
unless the ontology explicitly entails %
the contrary.
In this scenario, the
decision criterion eliminates all unfounded set checks. However,
as there is only one compatible set per instance, there would be only one
unfounded set check anyway, hence the speedup due to the decision criterion
is not significant.
But the effect of the decision criterion
can be increased by slightly modifying the scenario such that there
are multiple compatible sets. This can be done, for instance, by
nondeterministic default classifications, e.g., if a wine is not
Italian, then it is either French or Spanish by default.
Our experiments have shown that with a small number of compatible sets,
the performance enhancement due to the decision criterion is marginal,
but increases with larger numbers of compatible sets. For instance,
for $243$ compatible sets (and thus $243$ unfounded set checks)
we could observe a speedup from $13.59$ to $12.19$
seconds.

\section{Conclusion}
\label{sec:conclusion}

The evaluation of \hex-programs requires a minimality check of model
candidates which is realized as an equivalent search for unfounded sets (UFS).
However, this check is computationally costly. Moreover, during construction
of the model candidate, the ASP solver used as a backend has already performed
a ``restricted'' form of unfounded set check, i.e., an UFS check over
the program $\ProgramP$, viewing external atoms as ordinary ones. Hence,
it already excludes certain unfounded candidates. Redoing a complete UFS
search is thus a waste of resources, and the goal is to
minimize the number of additional foundedness checks.

In this paper we presented a syntactic criterion which can be
efficiently tested and allows to decide whether an additional UFS check
is necessary for a given program. It turned out
that %
the essential property is the existence of cyclic dependencies of atoms
which involve predicate inputs to external atoms. If no such
dependencies exist, then there is no need for an additional check, and
the check built into the ordinary ASP solver is already sufficient. In
further elaboration, we have refined the basic idea by splitting the
input program into components. This allows for independent applications
of the decision criterion to the different components. Thus, the UFS
check is restricted to relevant parts of the program, while it can
safely be ignored for other parts.

Related to our work is~\cite{Drescher08conflict-drivendisjunctive}, where
a similar program decomposition is used, yet
for ordinary programs only. While we consider e-cycles, which are specific
for \hex-programs,
the interest in \cite{Drescher08conflict-drivendisjunctive} is with head-cycles
with respect to disjunctive rule heads. In fact, our implementation
may be regarded as an extension of the work in~\cite{Drescher08conflict-drivendisjunctive},
since the evaluation of $\ProgramP$ follows their principles of performing UFS checks
in case of head-cycles.
Note however, that the applied component splitting %
does not generalize the well-known splitting theorem~\cite{lif94}
as we consider only positive dependencies for ordinary atoms.

An interesting issue for further research is to consider refinements
of the decision criterion, or alternative criteria. One direction for
refinement is to dynamically take the model candidate into account, in
addition to the program structure, which intuitively may prune
dependencies and thus allow to skip the UFS check even in the presence
of (syntactic) e-cycles. Another extension is to exploit additional
semantic information on the external atoms, e.g., such as
(anti-)monotonicity etc. Moreover, a more extensive experimental
analysis is subject of our future work, where case studies may give rise to
to alternative criteria and further optimizations.

\ifinlineref

\else
\bibliographystyle{splncs03}
\bibliography{unfoundedsetcheck}
\fi

\end{document}